\documentclass[11pt, oneside]{article}
\usepackage[T1]{fontenc}
\usepackage[english]{babel}
\usepackage[a4paper, margin=1in]{geometry}
\usepackage{setspace}
\usepackage{parskip}
\usepackage{fancyhdr}
\usepackage{enumitem}
\usepackage{array}
\usepackage{booktabs}

\usepackage{caption}
\usepackage{pdflscape}
\usepackage{longtable}
\usepackage{tabularx}
\usepackage{multirow}
\usepackage{tikz}
\usetikzlibrary{arrows.meta,decorations.pathmorphing,positioning, calc}
\usepackage{microtype} 
\usepackage{graphicx}
\usepackage{float}
\usepackage{caption}
\usepackage{pgf-pie}
\usepackage{multirow}
\usepackage{ragged2e}
\usepackage{sectsty}
\usepackage{pdfpages}
\usepackage{hyperref}
\usepackage[style=apa, backend=biber]{biblatex} 
\title{Requirements for a cooperative information infrastructure for the digital preservation of scholarly blogs}
\date{}

\begin{document}

\author{
Catharina Ochsner\thanks{Berlin School of Library and Information Science, Humboldt-Universität zu Berlin, Dorotheenstr.~26, 10117 Berlin, Germany.
Email: \href{mailto:catharina.ochsner@hu-berlin.de}{catharina.ochsner@hu-berlin.de}. \textbf{ORCID}: \href{https://orcid.org/0009-0005-3885-3951}{0009-0005-3885-3951}}\\
{\small HU Berlin}
 \and 
Heinz Pampel\thanks{Berlin School of Library and Information Science, Humboldt-Universität zu Berlin, Dorotheenstr.~26, 10117 Berlin, Germany und Helmholtz Association, Helmholtz Open Science Office, Telegrafenberg, 14473 Potsdam, Germany.
Email: \href{mailto:heinz.pampel@hu.berlin.de}{heinz.pampel@hu-berlin.de}. \textbf{ORCID}: \href{https://orcid.org/0000-0003-3334-2771}{0000-0003-3334-2771}}\\
{\small HU Berlin \& Helmholtz Association}
}

\maketitle

\pagestyle{fancy}
\fancyhf{}
\fancyhead[R]{\thepage}
\section{Abstract}
The long-term accessibility and reusability of scholarly knowledge is a central concern of Open Science. Research and infrastructure development in this area have so far focused predominantly on how traditional scientific outputs, such as journal articles, monographs, and conference proceedings, can be preserved and made openly available over time. Alternative forms of scholarly communication such as scholarly blogs, by contrast, have received comparatively little attention, even though they have become an established medium for disseminating research and for fostering dialogue within academia and with the wider public. The lack of preservation of scholarly blogs puts them at a risk of information loss, which poses the threat of leaving a gap in the scholarly record. Prior research has examined how blogs are integrated into information infrastructures and what requirements scholarly bloggers have for an information infrastructure that ensures long-term access to their blogs. What is needed now are recommendations for the implementation of these results into the library practice. Based on a convergent mixed-methods design that merges a quantitative analysis of 866 German scholarly blogs, a qualitative interview study with 13 scholarly bloggers, and an open, participatory review process with the scholarly blogging community, we propose a catalog of requirements for the integration of scholarly blogs into information infrastructures in order to ensure their long-term accessibility, reusability and citeability.
\newpage

\pagenumbering{arabic}

\section{Introduction}
Scholarly blogs constitute an important component of the scholarly communication culture in between scholars and also beyond the academic community such as journalists and the wider public \autocite{luzon2013}. Blogs are written by scholars, university or college students, teachers, science journalists, or practitioners who blog about research or research related topics \autocite{blanchard2013, bonetta2007, goldstein2009, kjellberg2015} and can do so rapidly and openly, to stimulate discussion, and to foster dialogue \autocite{burton2015, fenner2022}. In contrast to more traditional publication formats, such as journal articles or monographs, scholarly blogs face particular challenges in respect to their digital preservation and long-term accessibility. Blogs can be lost due to technical changes and the absence of preservation strategies \autocite{ochsner2025_jdoc}. Even though some scholarly bloggers, platform providers, and information infrastructure institutions have made isolated efforts to safeguard scholarly blogs \autocite{ochsner2026_iconf, ochsner2025_jdoc}, there remains a lack of standardized procedures for making blog posts permanently discoverable and unambiguously citable \autocite{ochsner2025_jdoc}.

In prior work we have examined to what extent scholarly blogs are integrated into digital research and information infrastructures \autocite{ochsner2025_jdoc, ochsner2025_bfp}, how bloggers describe their preservation practices \autocite{ochsner2026_iconf}, and what bloggers requirements for an information infrastructure for scholarly blogs are \autocite{ochsner2026_requirements_preprint}. In this article we propose solutions for the practical implementation of such an infrastructure. The work was guided by the following research questions (RQs):

\begin{table}[H]   \captionsetup{justification=centering, format=plain}
    \centering
    \renewcommand{\arraystretch}{1.1}
    \begin{tabular}{p{1.0\linewidth}}
    \textbf{RQ1:} What are the requirements for an information infrastructure for the digital preservation of scholarly blogs in order to ensure their long-term access?\\
    \hspace{2em}\textbf{RQ 1.1.} What are the functional requirements?\\
    \hspace{2em}\textbf{RQ 1.2.} What are the requirements concerning data protection of blogs and bloggers?\\
    \hspace{2em}\textbf{RQ 1.3.} What are the requirements for the aggregation of blogs?\\
    \hspace{2em}\textbf{RQ 1.4.} What are the requirements for the import and export of data?\\
    \end{tabular}
\end{table}

To answer these questions we developed a catalog of requirements for the integration of scholarly blogs into information infrastructures. The catalog defines requirements designed to ensure that scholarly blogs can persist as citable contributions to scholarly communication. This work is relevant for scholarly bloggers, scholarly blogging platforms providers, and information infrastructure organizations such as libraries, repositories, and archives that seek to ensure the long-term preservation of blogs. The defined requirements serve as a foundation for the development and evaluation of technical workflows as well as the establishment of archiving strategies. In doing so, the requirements catalog aims to contribute to sustainability and reproducibility of the scholarly record.
\section{Background}
Blogs (short for weblogs) are publicly accessible webpages on which authors publish diary- or journal-style content in reverse chronological order \autocite{nentwich2012}. Blogs first emerged in the late 1990s \autocite{rettberg2008} and gained prominence with the emergence of the Web 2.0 in the 2000s \autocite{fenner2008}, which marked a shift toward participatory digital environments characterized by increasingly user-generated content and greater opportunities for bottom-up contributions to online communication \autocite{nentwich2012}. This shift also created new opportunities for scholarly communication, as the internet increasingly became a space for disseminating research \autocite{nentwich2003, nentwich2012}. Ensuring the accessibility of web content that is used for research purposes is necessary in order to act according to good scientific practice \autocite{allea2023}. Long-term accessibility of web content can be achieved through digital preservation, a wide range of activities that enable the long-term access to digital and traditional cultural heritage collections \autocite{ala_preservation_statistics}, which is primarily enacted through libraries and archives but also museums and historical societies \autocite{ala_preservation_statistics, borgman2007, nentwich2003}. Digital preservation is often enacted through web archiving \autocite{lee2002}, with the archival of blogs being a part of web archiving \autocite{kasioumis2014}. In the following section we will list actors relevant to blog preservation and describe existing practices to ensure the long-term accessibility of blogs.

\subsection{Actors in blog preservation}
The actors described in the following are drawn from the German context and represent categories — web archives, dedicated blog archives, national libraries, subject-specific information services, institutional repositories, union catalogs and periodicals databases, and scholarly blogging platform providers, — that exist in analogous forms in many national information infrastructures.

\paragraph{Web Archives} Digital preservation can be enacted through web archiving \autocite{lee2002}, which plays a significant role in safeguarding cultural heritage and knowledge \autocite{kalb2013}. Web archives typically operate either as broad, non-selective crawls of the open web or as curated collections built in cooperation with libraries and other memory institutions. The most widely used general-purpose web archive is the Internet Archive, a non-profit initiative that provides a digital library of web pages \autocite{internet_archive_nodate, kalb2013}. Through the Internet Archive's Wayback Machine, over 29 years of web history are accessible. The Internet Archive collaborates with more than 1,250 libraries and other partners that identify and preserve relevant web pages within the Archive-It program. Furthermore, private individuals are also able to save web pages via the Internet Archive \autocite{internet_archive_nodate}.

\paragraph{Dedicated Blog Archives} Dedicated blog archives are services that specialize in the systematic ingestion, description, and preservation of scholarly blogs. Unlike general-purpose web archives, they combine blog-specific harvesting with scholarly infrastructure components, including persistent identifiers, metadata standards, and open licensing. Efforts towards blog archiving have been made within the EU-funded BlogForever project. The repository produced by the project is, however, no longer available \autocite{pampel2024}. A prominent example for a blog archive is Rogue Scholar, an InvenioRDM-based platform launched in 2023 and operated by Front Matter \autocite{fenner2022, fenner2023}. As of October 2025, Rogue Scholar lists 171 archived blogs \autocite{rogue_scholar_nodate}. The platform archives blog content under a Creative Commons Attribution (CC BY) license \autocite{hofting2025} and additionally ensures web archiving via the Internet Archive since Rogue Scholar holds a subscription to the Internet Archive's Archive-It service, which enables the long-term archiving of all blogs registered with Rogue Scholar \autocite{rogue_scholar_nodate}. Rogue Scholar also assigns Digital Object Identifiers (DOIs) to individual posts \autocite{fenner2022}. A DOI is a unique, permanent, and distinct digital identifier for an object that enables reliable access to that object \autocite{doi_nodate}. DOIs are enriched with metadata, which enables the reuse for research or other services \autocite{strecker2023, van_eck2025}.

\paragraph{National Libraries} National libraries typically hold the mandate for the long-term collection, cataloging, and preservation of a country's published output, and many have extended this mandate to selected categories of online publications. Their role is particularly relevant for scholarly blogs insofar as national libraries can provide persistent identifiers such as International Standard Serial Numbers (ISSNs), which are an identification system for serial publications that in Germany, are often distributed through the German National Library \autocite{dnb_issn}. National libraries can furthermore provide authoritative catalog records, and long-term digital preservation services. For example, the German National Library has collected online publications since 2006. A comprehensive collection of web pages through manual review is, however, not feasible due to limited financial resources. Access to collected online publications and blogs is generally only available in the reading rooms of the German National Library, as the necessary licenses for online access are lacking \autocite{weisbrod2010}. Long-term digital preservation is carried out through the archiving system Koala, in which digital objects and their metadata are archived and subsequently imported into the library's catalog \autocite{dnb2019}. The German National Library does not distinguish between scholarly and non-scholarly blogs in its collection and archiving activities. As of August 2024, 289 scholarly blogs were listed in the German National Library's catalog \autocite{ochsner2025_oat}. In addition, the German National Library assigns ISSNs to blogs held in its collections \autocite{dnb_issn, ochsner2025_jdoc}. Apart from the German National Library, blogs already have entries in the catalogs of other national libraries, such as the National Library of Australia \footnote{https://www.library.gov.au/}, the Bibliothèque nationale de France\footnote{https://www.bnf.fr/}, the National and University Library in Zagreb\footnote{https://nsk.hr/en/}, or the Bibliothèque et Archives Canada\footnote{https://bac-lac.on.worldcat.org/discovery}.

\paragraph{Subject-Specific Information Services} Many national research systems maintain subject-specific information services that provide supra-regional access to specialized scholarly literature and, in some cases, assume preservation responsibilities for the publications they acquire. Such services bridge the gap between institutional repositories and national-level preservation by offering discipline-oriented collections and workflows. In Germany, this function is performed by the Specialized Information Services (ger: Fachinformationsdienste, FID), which are typically anchored at scholarly institutions and libraries. All print and electronic media acquired through the Specialized Information Services must be permanently archived by the responsible libraries in accordance with the mandate of the German Research Foundation (ger: Deutsche Forschungsgemeinschaft, DFG). Specialized Information Services also assume responsibilities in the context of open access and make publications available on repositories \autocite{wirtz2024}.

\paragraph{Institutional Repositories} Institutional repositories are maintained by universities, research institutes, and other scholarly organizations to collect, describe, and make accessible the scholarly output of their members. While they have traditionally focused on journal articles, theses, and research data, some institutions have extended their scope to include blog posts authored by affiliated researchers, thereby integrating blogs into the same preservation and discovery workflows as other scholarly outputs. Examples include the Max Planck Institute for Legal History and Legal Theory\footnote{https://www.lhlt.mpg.de/en}, the Max Planck Institute for the Study of Religious and Ethnic Diversity\footnote{https://www.mmg.mpg.de/home/}, and the Max Planck Institute for the Study of Societies\footnote{https://www.mpifg.de/2733/en} which index and provide access to blog posts through the Max Planck Society repository PURE. A further example is the Potsdam Institute for Climate Impact Research\footnote{https://www.pik-potsdam.de/en}, which makes blog posts by its researchers available — including, for instance, selected contributions by Stefan Rahmstorf's blog Klimalounge\footnote{https://scilogs.spektrum.de/klimalounge/}.

\paragraph{Union Catalogs and Periodicals Databases} Union catalogs and periodicals databases aggregate bibliographic records for serially published works across multiple institutions and countries, thereby supporting discovery, interlibrary lending, and the coordinated assignment of identifiers such as ISSNs. They typically describe publications at the bibliographic level rather than archiving their full content. For the German-speaking area, this function is taken up by the Periodicals Database (ger: Zeitschriftendatenbank, ZDB), which records all serially published print and online publications held in German and Austrian institutions and continuously expands European access to digitized newspapers \autocite{zeitschriftendatenbank_nodate}. A search for the term "blog" yields approximately 265 results (as of October 16, 2025). The catalog entries, however, capture exclusively bibliographic information, including ISSNs, while the content of the blogs themselves is not archived.

\paragraph{Platform Providers} Scholarly blogging platforms provide shared hosting, editorial, and visibility services for academic blogs, often with a disciplinary focus. By centralizing technical maintenance and basic metadata workflows, they lower the threshold for scholars to publish blogs and at the same time enable coordinated preservation and citation practices across many individual blogs. A prominent European example for the humanities and social sciences is Hypotheses. Hypotheses is part of OpenEdition, a comprehensive digital publishing service for the dissemination of works in the humanities and social sciences, offering blogs in a wide variety of languages. All blogs and texts are published open access \autocite{enhypotheses_nodate}. The German-language portal de.hypotheses is operated by OpenEdition in collaboration with the Max Weber Foundation \autocite{dehypotheses_nodate}. OpenEdition is a portal for electronic resources in the humanities and social sciences \autocite{openedition_nodate}. OpenEdition is part of the Centre pour l'édition électronique ouverte (CLEO, Centre for Open Electronic Publishing), which is affiliated with the École des Hautes Études en Sciences Sociales (EHESS) \autocite{ehess_2015}, a French public research institution \autocite{ehess_2016}. Hypotheses provides DOIs and citation suggestions for blog posts and also coordinates ISSN assignment for blogs through the German National Library. Hypotheses blogs that receive an ISSN from the German National Library are also assigned a catalog entry and are archived for long-term preservation by the German National Library. The use of DOIs and citation suggestions is rarely adopted outside of de.hypotheses \autocite{ochsner2025_jdoc, ochsner2025_bfp}.
\section{Method}
To translate empirical evidence about scholarly blogs into actionable infrastructure requirements, we combined three complementary strands of inquiry: a quantitative analysis of 866 German scholarly blogs, a qualitative interview study with 13 bloggers, and an open, participatory review process with the wider scholarly blogging community in Germany. The purely quantitative mapping revealed that gaps exist in the integration of scholarly blogs into information infrastructures, but not how they should be closed; the purely qualitative study has surfaced bloggers' requirements, but without a baseline against which to assess their prevalence. We therefore adopted a convergent mixed-methods design \autocite{creswell2017}, in which the quantitative and qualitative strands were collected and analyzed separately and subsequently merged through integrative data analysis (see Figure \ref{fig:1}). We extended this design with an open, participatory review process, which we treat as a communicative validation layer, since requirements that survive both empirical strands and the scrutiny of practitioners and infrastructure experts are more robust than those derived from a single source. While the underlying empirical studies have been reported in detail elsewhere \autocite{ochsner2025_jdoc, ochsner2026_requirements_preprint, ochsner2026_iconf, ochsner2025_bfp}, the contribution of this article lies in the integration of their findings into a coherent, implementation-oriented catalog of requirements and in its validation through community review. In the remainder of this section, we summarize the research designs of the two empirical studies, before describing the merging step and the participatory review process in greater detail.

\subsection{Study 1: Quantitative Data Analysis}
For the quantitative study we curated a sample of 866 German scholarly blogs. We started off by performing a web search and identifying sources that already compiled recommendations or lists of scholarly blogs. We found lists of blogs in the context of the annual prize for the Scholarly Blog of the Year (ger: Wissenschaftsblog des Jahres) by the German blog Science communicating (ger: Wissenschaft kommuniziert) that took place from 2011-2022 \autocite{wissenschaftkommuniziert2022}. We then included blogs mentioned in the project proposal Infra Wiss Blogs \autocite{pampel2024}, we collected blogs of the German research performing institutions Jülich Center and the Helmholtz Association, and included blogs from the State Library Berlin. We then included blogs from the blogging platforms de.hypotheses (using the affiliated catalog OpenEdition), ScienceBlogs and SciLogs. Lastly we included scholarly blogs cataloged by the German National Library, included blogs we discovered through desk research, used the digital blog archive Rogue Scholar, and went through the blogrolls of all the blogs in our sample. We have summarized these steps and the amount of blogs we collected in each step in Table \ref{table:1}.

\begin{table}[H]
    \captionsetup{justification=centering, format=plain}
    \caption{Blog collection process}
    \label{table:1}
    \centering
    \renewcommand{\arraystretch}{1.0}
    \begin{tabular}{p{0.45\linewidth} p{0.45\linewidth}}
        \toprule
        \textbf{Step} & \textbf{Outcome} \\
        \midrule
        Nominees of the "Scholarly Blog of the Year" (2011–2022) & 34 blogs \\
        Project proposal & 7 blogs \\
        Helmholtz Association blogs & 15 blogs \\
        Jülich Center blogs & 16 blogs \\
        State library Berlin blogs & 10 blogs \\
        Open Edition & 446 blogs \\
        ScienceBlogs & 88 blogs \\
        SciLogs & 24 blogs \\
        German National Library blog entries & 76 blogs \\
        Desk research & 41 blogs \\
        Rogue Scholar & 14 blogs \\
        Blogroll & 95 blogs \\
        \midrule
        \textbf{Final sample size} & \textbf{866 blogs} \\
        \bottomrule
    \end{tabular}
\end{table}

Following the sampling of 866 German scholarly blogs, we developed three sets of analysis criteria. The first set of analysis criteria was aimed at collecting data that would provide an overview over the German scholarly blogging landscape by examining blog activity, institutional affiliation, discipline, and language. The second set was aimed at collecting data in order to assess how scholarly blogs and their content are integrated into existing digital research and information infrastructures by examining platform and software choices, checking whether the blogs are indexed by the German National Library, and whether or not blogs had registered an ISSN on the blog level, and DOIs and citation proposals for blog posts. The last set aimed at collecting data on strategies bloggers apply to enhance the accessibility and reusability of their content by searching for blog archives, blogrolls, Creative Commons (CC) licenses, and blog feeds. Finally, we collected the necessary data via autopsy for each blog individually and quantitatively analyzed the data using R. The complete data set is available on Zenodo \autocite{ochsner2025_dataset}.

\subsection{Study 2: Qualitative Data Analysis}
For the second, qualitative study, we adopted an exploratory qualitative design based on semi-structured interviews, using an interview guide for structure that also allowed for flexibility in phrasing and question order, dependent on each interview. This allowed us to maintain flexibility in the interaction with participants. We started each interview with general questions about the participants and their background, their blogging activities and their history. We continued with more specific questions on infrastructure and preservation, bloggers experiences and their requirements. We intentionally did not define the terms “information infrastructure”, “preservation”, or “long-term accessibility” and let participants articulate their own interpretation of these terms to foreground participants' perspectives.

Table \ref{table:2} provides an overview of the participants, their discipline, their gender as perceived by the authors, and their career stage according to the European Commission's Researchers’ Career Framework which classifies research careers into the following four stages: First Stage Researchers (R1), Recognized Researcher (R2), Established Researcher (R3), and Leading Researcher (R4) \autocite{european_commission2011}. The blogging platform of individual participants was omitted to ensure the participants’ anonymity.

\begin{table}[H]
    \captionsetup{justification=centering, format=plain}
    \caption{Participants}
    \label{table:2}
    \centering
    \renewcommand{\arraystretch}{1.1}
    \begin{tabular}{p{0.17\linewidth} p{0.3\linewidth} p{0.2\linewidth} p{0.2\linewidth}}
        \toprule
        \textbf{ID} & \textbf{Discipline} & \textbf{Perceived gender} & \textbf{Career stage}\\
        \midrule
        P1 & Humanities \& Social Sciences & Male & R1\vspace{2pt}\\
        P2 & Humanities \& Social Sciences & Male & R2\vspace{2pt}\\
        P3 & Humanities \& Social Sciences & Male & R4\vspace{2pt}\\
        P4 & Humanities \& Social Sciences & Female & R1\vspace{2pt} \\
        P5 & Natural Sciences & Male & R1\vspace{2pt}\\
        P6 & Life Sciences & Female & R1\vspace{2pt}\\
        P7 & Humanities \& Social Sciences & Female & R1\vspace{2pt}\\
        P8 & Natural Sciences & Male & R2\vspace{2pt}\\
        P9 & Humanities \& Social Sciences & Female & R3\vspace{2pt}\\
        P10 & Humanities \& Social Sciences & Male & R1\vspace{2pt}\\
        P11 & Engineering Sciences & Male & R1\vspace{2pt}\\
        P12 & Humanities \& Social Sciences & Female & R2\vspace{2pt}\\
        P13 & Humanities \& Social Sciences & Male & R4\vspace{2pt}\\
        \bottomrule
    \end{tabular}
\end{table}

We aimed at sampling a diverse range of participants concerning blogging platform, discipline, gender identity, and research stage. We ended up conducting 13 interviews and two pre-tests. The pre-tests showed that participants found it challenging to name preservation practices. We therefore removed questions that were too general and provided the participants with examples and scenarios, they could relate to. We contacted participants via e-mail and provided them with information about the study, the procedure and our data protection measures. All participants consented to the conduct, the recording and the transcription of the interviews. We conducted the interviews in German between January 30 and March 26, 2025 and recorded them via Zoom \footnote{https://zoom.us/de/join} and one via the open-source platform BigBlueButton\footnote{https://bigbluebutton.org/}. The interviews ranged between 35 and 51 minutes. To ensure anonymity, neither the audio recordings nor the transcripts have been made publicly available and each participant was assigned a participant ID (P1–P13).

We transcribed the interviews using the GDPR-compliant software Amberscript\footnote{https://www.amberscript.com/de/} and reviewed each transcript to ensure accuracy and completeness. We analyzed the data according to qualitative content analysis approach by Kuckartz and Rädiker (\cite{kuckartz2024}). The transcripts were imported into the qualitative research software MAXQDA and coded using a structured coding systems, made out of deductive codes that we derived from the interview guide and inductive codes that emerged from the data. The coding frame was organized hierarchically across multiple levels.

\subsection{Study 3: Merging the Results}
\begin{figure}[H]
  \centering
    \begin{tikzpicture}[
    roundnode/.style={circle, draw=blue!20, fill=blue!5, thick, minimum size=1cm, align=center},
    squarednode/.style={rectangle, draw=blue!20, fill=blue!5, thick, minimum height=1.2cm, text width=3cm, align=center}]
    \node[roundnode](quanti){Study 1: \\ Quantitativ \\ Data Analysis};
    \node[roundnode](quali)[below=of quanti]{Study 2: \\ Qualitative \\ Data Analysis};
    \node[squarednode](results)at($(quanti)!0.5!(quali)$) [right=2cm] {Merging the\\ Results};
    \node[squarednode](rfc)at($(quanti)!0.5!(quali)$) [right=6cm] {Request for\\ comments};
    \node[roundnode](interpret)[right=of rfc] {Study 3:\\ Interpret Results\\ to Derive Catalog};
    \draw[->] (quanti) -- (results);
    \draw[->] (quali) -- (results);
    \draw[->] (results) -- (rfc);
    \draw[->] (rfc) -- (interpret);
    \end{tikzpicture}
    \caption{The Convergent Mixed Methods Design (One-Phase Design) according to \cite{creswell2017}, expanded to include the open peer review process}
  \label{fig:1}
\end{figure}
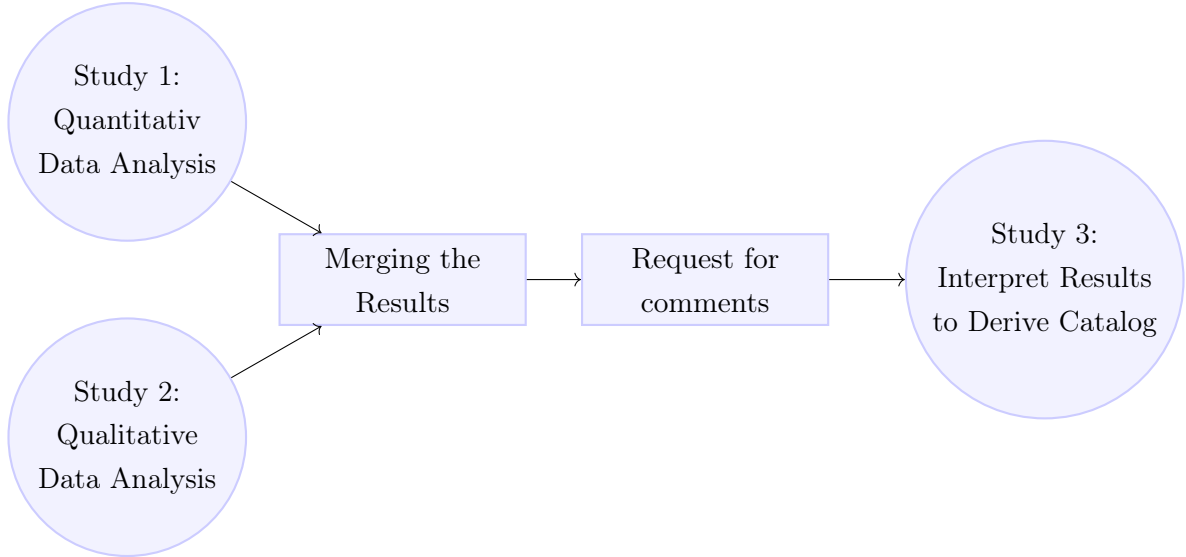

Lastly, we compared the quantitative and qualitative data by merging their results using integrative data analysis. The quantitative and qualitative data were collected and analyzed separately \autocite{creswell2017}. While the quantitative work identified gaps in the integration of scholarly blogs into digital research and information infrastructures the data we collected in the interview study identified possible solutions to fill this gap. We therefore applied a convergent mixed methods design that merged quantitative and qualitative results, as shown in Figure \ref{fig:1} \autocite{creswell2017}. We furthermore expanded on the convergent mixed method design by publishing a draft of our findings in a request for comments phase in order to include a participatory peer review process that allowed experts from information infrastructure facilities and bloggers to give feedback on this work.

Finally, this study employed an open, collaborative review process, in which feedback was solicited from peers via a shared document platform. A draft version of the catalog was shared via a collaborative document, where peers were invited to provide comments using the platform’s commenting and suggestion features. The draft was published in German on January 12th 2026. We closed the commenting phase on February 27th, 2026. We received 22 comments, which mostly commented on the content and suggested additional content we were lacking. The feedback was incorporated into a subsequent revision. This process did not follow a formal peer-review protocol but served as an open, collaborative review stage to improve clarity and content. The whole draft was then translated into English by the authors. Additionally, we hosted and moderated a virtual discussion, in which we presented the catalog, discussed it with 40 participants and included their feedback into the final draft.

\section{Findings}
\subsection{Study 1: Quantitative Data Analysis}
In our quantitative study we showed that even though the German blogosphere is diverse, it is distinctly shaped by the humanities and social sciences. The scholarly blogging platform de.hypotheses has contributed substantially to the dissemination and visibility of blogs in these fields, while the natural sciences, life sciences, and engineering sciences are considerably less represented, in part because comparable platforms in these areas are lacking. Furthermore, the Social Sciences and Humanities are still predominantly represented even without considering blogs hosted by de.hypotheses \autocite{ochsner2025_jdoc}. The predominance of the Social Sciences and Humanities is furthermore supported by another dataset of German scholarly blogs \autocite{gesis, voigt2024}. Although many German blogs publish in German, a relevant proportion (approximately one quarter) blog entirely or partly in English \autocite{ochsner2025_jdoc}. Scholarly blogs are only partially cataloged or archived for long-term preservation by the German National Library. Likewise, the assignment of ISSNs or DOIs remains rare. Efforts are made by de.hypotheses, which regularly submits blogs to the German National Librarx for the assignment of ISSNs \autocite{dehypotheses_FAQ}. German scholarly blogs that predominantly use DOIs are blogs hosted by de.hypotheses, who assign DOIs provided by DataCite and encourages bloggers to provide citation proposals \autocite{dehypotheses_FAQ}. Although de.hypotheses is used by many bloggers in the Social Sciences and Humanities, some bloggers also blog on ScienceBlogs, SciLogs or Wordpress \autocite{ochsner2025_jdoc}. Open licenses are also frequently absent, which complicates the reuse of blog posts. The majority of blogs use software provided by WordPress\footnote{https://wordpress.org/download/}. Majority of blogs use feeds such as Atom, JSON or RSS, which bring the potential of integration into other systems \autocite{ochsner2025_jdoc}. 

\subsection{Study 2: Qualitative Data Analysis}
The findings of the above described quantitative study match bloggers' statements on their preservation practices to a high extent. Participants recognized the importance of long-term preservation of scholarly blogs but expressed a strong desire for control over what content gets preserved. When asked how participants enact long-term accessibility and reuse, bloggers stated to receive DOIs and ISSNs, mostly through de.hypotheses. Some blogs had a catalog entry in an academic or a national library. However, some participants also stated to use personal preservation practices to secure blog content, such as local backups, reusing blog content in other publication formats, like monographs, presentations or newsletter, using CC licenses or the Internet Archive \autocite{ochsner2026_iconf}. Bloggers require a long-term, accessible, and trustworthy infrastructure for preserving their content but disagree about what kind of content is preservation worthy. Some participants argue for archiving everything if storage allows, others say only selected posts matter, while most participants want bloggers to control what content gets preserved. Commonly identified items for preservation include texts, images, comments, audio, and links, while material such as calls for papers, outdated or non‑scientific posts, are often considered not necessary. Technically and organizationally, participants emphasized reliable long‑term access: stable URLs, citability (DOIs), searchable archives (e.g., by date), and transparency about access conditions. Desired services ranged from editorial support (updating posts, peer review, checking affiliations, filtering irrelevant content) to safeguards against retrospective alteration. While participants agreed that individual bloggers cannot be solely responsible they also accepted that bloggers must prepare content for preservation through open licenses or metadata. However, metadata and metadata literacy were highlighted as essential, with calls for automation and tools to lower the burden on bloggers who lack technical expertise. Several participants favored government‑funded institutions like national or university libraries as primary custodians, expressing distrust of private actors and concern about politicized platforms. Others urged a mixed, international governance model to reduce single‑nation or corporate control and to build resilience for politically sensitive scholarship. Digital archives such as the Internet Archive were praised as valuable but also seen as unstable due to sustainability risks, and governance concerns tied to US‑based foundations. Blogging platforms (e.g., Hypotheses, SciLogs, ScienceBlogs) are commonly relied on by bloggers. In sum, durable preservation of scholarly blogs requires aligning intended future uses with appropriate capture strategies, combining technical features (stable identifiers, metadata, searchability), organizational services (transparency, automation), and trustworthy governance (public and decentralized). Practical solutions should mix institutional infrastructures, platform cooperation, automated tooling, and blogger actions to ensure accessible, trustworthy, and resilient long‑term preservation \autocite{ochsner2026_requirements_preprint}.

\subsection{Study 3: Merging the Results}
The findings show that while individual bloggers apply strategies that facilitate reuse, the long-term visibility and accessibility of many scholarly blogs remains uncertain. Although various actors and institutions already process scholarly blogs as objects of archiving and cataloging, there is currently no unified metadata schema nor a comprehensive initiative to make scholarly blogs persistently accessible and reusable \autocite{ochsner2025_jdoc}. In the next section we combine the findings of the quantitative and qualitative study. While the quantitative study uncovered gaps in the integration of scholarly blogs into digital research and information infrastructures, the findings of the qualitative study fill these gaps. We articulate gaps and their solution in the requirements for an information infrastructure below. We will answer each RQ and start with defining functional requirements, followed by requirements concerning data protection, define requirements for the aggregation of blogs and lastly describe requirements for the import and export of data.

\subsubsection{Functional}

\paragraph{F-01: Assignment of Persistent Identifiers such as Digital Object Identifiers (DOIs) at the Blog Post Level and ISSNs at the Blog Level}\par

Blogging platforms / repositories ensure that blogs are addressable via ISSNs and that all posts are addressable via a DOI, which have become the central persistent identifier (PID) system for scholarly content in recent years. DOIs are assigned, for example, through Crossref\footnote{https://www.crossref.org/} or through DataCite\footnote{https://datacite.org/} for a wide range of research-related scholarly objects \autocite{crossref, vierkant_datacite}. For self-registration, blog operators incur costs such as membership fees and individual DOI registration charges. Platforms such as de.hypotheses and Rogue Scholar cover these costs through framework agreements with DOI registration agencies \autocite{dehypotheses_FAQ, fenner_doi_2025}.

The aim is to provide persistent addressing of blogs and blog posts at the level of individual contributions. ISSNs may be requested from libraries such as national libraries, by bloggers. Bloggers using de.hypotheses can request an ISSN through their platform. Bloggers can have DOIs assigned, for example, through Rogue Scholar or de.hypotheses. All blog posts archived by Rogue Scholar carry a DOI registered with either Crossref or DataCite, along with relevant metadata. For blogs for which Rogue Scholar manages DOI registration, DOIs are automatically registered for each new blog post. DOIs are also assigned through de.hypotheses. Blogs are eligible once they are listed in the OpenEdition catalog. Each newly published blog post then automatically receives a DOI, which is also included in the metadata. In addition, each post receives its own document page on the OpenEdition website. Furthermore, other information infrastructure services also assign DOIs via repositories (see F-03).

\paragraph{F-02: Metadata}
Bloggers provide comprehensive metadata for their blog posts. Metadata such as language, title, and abstract of a blog post, as well as the names of authors, support the discoverability of blog posts.

The aim is to enhance the citability and discoverability of blog posts.
Table 2 \ref{table:3} lists a set of metadata that is required and further metadata that is recommended.

\begin{table}[H]
    \captionsetup{justification=centering, format=plain}
    \caption{Required and recommended Metadata}
    \label{table:3}
    \centering
    \renewcommand{\arraystretch}{1.1}
    \begin{tabular}{p{0.15\linewidth} p{0.2\linewidth} p{0.55\linewidth}}
        \toprule
        \textbf{Property} & \textbf{Required} & \textbf{Description}\\
        \midrule
        Title & Yes & The title of the blog post. \vspace{3pt}\\
        Site Name & Yes & The title of the blog. \vspace{3pt}\\
        Authors & Yes & The authors of the blog post. \vspace{3pt}\\
        License & Yes & The license for the blog post. Rogue Scholar accepts \par only content published under a Creative Commons \par Attribution (CC-BY) license.\vspace{3pt}\\
        ORCID & Recommended & The ORCID of the authors and contributors of a post. \vspace{3pt} \\
        Description & Recommended & An abstract of the blog post.\vspace{3pt}\\
        Language & Recommended & The language in which the blog post is written.\vspace{3pt}\\
        Publication date & Recommended & The date of publication of the post.\vspace{3pt}\\
        Date of \par archiving & Recommended & The date on which the post was archived. \vspace{3pt}\\
        DOI or URL & Recommended & The URL or DOI of the post. \vspace{3pt}\\
        ISSN & Recommended & The ISSN of the blog (if available).\vspace{3pt}\\
        \bottomrule
    \end{tabular}
\end{table}

\paragraph{F-03: Integration of Scholarly Blogs into Repositories}
Repositories and other infrastructure services that index blogs ensure the integration of scholarly blogs and their posts into repositories and other services within the information infrastructure. This integration — in both subject-specific infrastructures such as Specialized Information Services (FIDs) and institutional infrastructures operated by universities and non-university research institutions — corresponds to the distributed character of the information infrastructure system in Germany but can also be applied to other national contexts.

The aim is to leverage repositories as a central infrastructure component for the sustainable archiving and reuse of scholarly blogs. In view of the distributed structure of the information infrastructure in Germany, a cooperative approach is required. Furthermore, a cooperative approach distributes responsibility and ensures the preservation of content by multiple actors. Rogue Scholar may occupy a potential role as an aggregator in this context. 

\paragraph{F-04: Content Representation of Blog Posts in Repositories}
Repositories and other infrastructure services that index bloggers ensure that the full-text rendering, the rendering of images, the correct display and reusability of links, and the integration of comments are enabled. In the archiving process, posts should be captured comprehensively (i.e., including texts, images, and comments) and in an automated fashion, so as to guarantee complete documentation. Through this practice, blogs are intentionally stored in multiple locations simultaneously in order to reduce the risk of information loss and to distribute responsibility. This requirement encompasses several sub-aspects:

\begin{itemize}
    \item \textbf{F-04.1: Full-text rendering}
    \item \textbf{F-04.2: Rendering of images}
    \item \textbf{F-04.3: Correct display and reusability of links}
    \item \textbf{F-04.4: Integration of comments}
\end{itemize}

The aim is the comprehensive preservation of blog posts. Harvesting via feeds (Atom, RSS, or JSON) offers significant potential in this regard, as conventional blogging software solutions generate these feeds in an automated manner.

\paragraph{F-05: References}
Bloggers support the referencing of cited literature. Bloggers who wish to cite external content either link to that content within their posts or compile a reference list.

The aim is to integrate blog posts as part of scholarly communication into the metadata ecosystem of scholarly publishing. A practice established by Rogue Scholar uses full-text content to extract references, employing the following strategies: if the blog uses citeproc (the most widely used library for formatting references, employed by reference managers such as Zotero, Mendeley, and many others) to generate reference lists; if reference sections begin with a heading (h1, h2, h3, or h4) bearing the label "References", followed by an ordered or unordered list. As a fallback strategy, Rogue Scholar searches for a heading labeled "References" and extracts all links found therein. Reference strings are not captured, as there are many different ways in which these may be formatted. Rogue Scholar then stores the references in a list with keys for each entry. This information is subsequently used to include the references in the DOI metadata registered with Crossref or DataCite and in the metadata registered with the Rogue Scholar repository software.

\paragraph{F-06: Versioning}
Bloggers support versioning. Changes such as substantial content revisions to results or discussion sections, editorially significant updates requiring a formal correction, changes made in response to peer reviews, and modifications to attached media files such as images or audio files must be designated as new versions.

The aim is to make the versioning of blog posts more transparent. Since feeds do not contain standardized metadata indicating the version of a blog post, feeds should be adapted by adding a link attribute that points to the previous version of the post \autocite{fenner_versioning_2025}.

\paragraph{F-07: Searchability}
Repositories and other infrastructure services that index blogs provide a searchable index that can be queried according to the attributes specified under F-02 Metadata.

The aim is to ensure that infrastructures preserving blogs are searchable by topic and by blog.

\subsubsection*{Data Protection}
\paragraph{D-01: GDPR Compliance}
Bloggers as well as repositories and other actors within the information infrastructure ensure that the General Data Protection Regulation (GDPR) serves as the governing framework for the secure and legally compliant processing of personal data.

The aim is for infrastructures to implement user consent mechanisms in order to process data lawfully. This includes informing users of their rights and ensuring the possibility of data deletion at any time.

\paragraph{D-02: Data Sovereignty}
Repositories and other actors within the information infrastructure ensure that the storage and processing of data is carried out in compliance with applicable data protection regulations.

The aim is for the selection of a hosting provider for scholarly blogs to involve servers located in the EU, in order to ensure GDPR compliance. Regular audits of data security practices are required to guarantee the security of stored data.

\subsubsection{Aggregation}

\paragraph{A-01: Opt-in}
Bloggers agree to have their content included in the infrastructure. This opt-in model respects the rights of bloggers and promotes engaged participation. An opt-out model, by contrast, would require bloggers to actively object, which could result in unintended archiving and should therefore be avoided \autocite{ochsner2025_oat}.

The aim is the registration of blogs in repositories at the initiative of and with the consent of bloggers. The registration and archiving of a blog in repositories follows these criteria: repositories consider only blogs that write about scholarly topics and whose full-text content is available via RSS feeds. A further condition is the distribution of blog content under the Creative Commons Attribution (CC BY) license.

\paragraph{A-02: Guidance and Support for Service Use}
Repositories and other actors within the information infrastructure ensure that all participating bloggers and institutions are able to use the infrastructure effectively. To this end, comprehensive guidance and support are indispensable. This encompasses training materials, workshops, and technical support in order to meet the needs of diverse user groups.

The aim is the provision of online tutorials, guides, and FAQs. In addition, regular webinars or workshops could be offered to familiarize bloggers with the benefits and possibilities of the infrastructure. A technical support service should also be established to provide immediate assistance with questions or issues.

\subsubsection{Interfaces}

\paragraph{I-01: Import Interfaces}
Bloggers provide blog posts via feeds (Atom, RSS, or JSON), thereby enabling, in accordance with the requirements described herein, the dissemination and preservation of blog posts in repositories and other infrastructure services.

The aim is the comprehensive preservation of blog posts by infrastructures such as repositories. Harvesting via feeds (Atom, RSS, or JSON) offers significant potential in this regard, as conventional blogging software solutions generate these feeds in an automated manner.

\paragraph{I-02: Export Interfaces}
Infrastructures such as repositories and other actors within the information infrastructure make stored blog posts available via interfaces such as OAI-PMH and REST APIs, so that they can be integrated into discovery services.

The aim is to enhance the discoverability of blog posts through discovery services. The provision of metadata and blog posts via standardized interfaces and protocols within the information infrastructure promotes the discoverability of blog posts and strengthens their visibility.
\section{Discussion}
In our previous work, we showed that the digital preservation of scholarly blogs requires a combination of technical requirements (persistent identifiers, standardized metadata, interoperable interfaces), organizational coordination between bloggers, platform providers, and infrastructure institutions, as well as legal compliance, in particular with respect to data protection. Taken together, the proposed requirements describe a cooperative information infrastructure rather than a single, centralized service. The principal contribution of this article is conceptual: building on our prior quantitative mapping of the German scholarly blogosphere \autocite{ochsner2025_jdoc, ochsner2025_bfp} and on our qualitative interview study of bloggers' requirements for an infrastructure for the preservation of scholarly blogs \autocite{ochsner2026_requirements_preprint}, we derive an implementation-oriented catalog of thirteen concrete requirements, structured along the four dimensions of functionality, data protection, aggregation, and interfaces. In doing so, we re-frame blog preservation from a question of individual institutional responsibility towards a model of distributed, cooperative stewardship. To our knowledge, no comparable work currently exists for scholarly blogs, and the requirements are, at the level of abstraction on which they are formulated, transferable to other national contexts. In the following, we discuss our findings along each RQ. We will begin by discussing our proposed functional requirements, followed by the requirements for data protection, the aggregation of blogs, and the requirements for the import and export of data.

\paragraph{Functional requirements (RQ 1.1)} The functional requirements (F-01 to F-07) deliberately mirror established practices in the digital research and information infrastructure for journal articles: persistent identifiers, structured metadata, repository integration, comprehensive content representation, referencing, versioning, and searchability. This alignment is a substantive claim: it positions scholarly blogs as a trustworthy and legitimate form of scholarly publication and of the scholarly record, rather than as ephemeral web content, while allowing bloggers and infrastructure providers to leverage existing services (Crossref, DataCite, OpenEdition, and libraries) rather than to construct parallel ones. At the same time, the requirements adapt these practices to the specifics of blogging through the distribution via feeds, the integration of multimedia and reader comments, fluid update cycles, and the frequent absence of formal editorial processes. Our empirical work shows that the technical components for these requirements are largely already in place; what is lacking is their systematic adoption across the heterogeneous blogosphere, particularly outside the humanities and social sciences \autocite{ochsner2025_jdoc}.

\paragraph{Data protection (RQ 1.2)} The data protection requirements (D-01, D-02) foreground a dimension that has traditionally received less attention in the preservation discussion but is particularly salient for scholarly blogs, based on bloggers' requirements for an information infrastructure \autocite{ochsner2026_requirements_preprint}. Unlike journal articles, blog posts frequently contain personal data of authors and commentators, include embedded third-party content, and are hosted on platforms (e.g. WordPress, Hypotheses, ScienceBlogs, SciLogs) whose data processing is governed by general privacy policies rather than by preservation-specific agreements. We propose that GDPR compliance must be designed into the preservation workflow from the outset rather than resolved case by case after the fact. This is also a prerequisite for bloggers' trust, which our interview study identified as a condition for voluntary participation \autocite{ochsner2026_requirements_preprint}.

\paragraph{Aggregation (RQ 1.3)} The aggregation requirements (A-01, A-02) reflect a deliberate design choice in favor of an opt-in rather than an opt-out model. A reviewer might object that opt-in limits coverage and therefore undermines the preservationist goal. We acknowledge this trade-off and nevertheless consider opt-in the appropriate default, for three reasons grounded in our empirical work: bloggers consistently emphasized the importance of remaining in control of what is archived \autocite{ochsner2026_requirements_preprint, ochsner2026_iconf}; scholarly blogs frequently contain content whose licensing status is ambiguous or whose authors regard certain posts as transient; lastly the coverage loss induced by opt-in can be mitigated by active outreach and low-threshold onboarding (A-02). A-01 and A-02 are therefore complementary: a participatory infrastructure requires both a respectful entry point and ongoing support, e.g. through tutorials, user support, and workshops \autocite{ochsner2026_requirements_preprint}.

\paragraph{Interfaces (RQ 1.4)} In the requirements for the import and export interfaces (I-01, I-02) we concretize the cooperative character of the proposed infrastructure. Import interfaces based on feeds (Atom, RSS, JSON) leverage a function that is already provided by the dominant blogging software, which will therefore minimize the burden on bloggers. Export interfaces via OAI-PMH and REST APIs connect the infrastructure to the broader ecosystem of discovery services, library catalogs, and subject-specific portals. This architectural decision draws an explicit lesson from the BlogForever project, whose repository is no longer available \autocite{pampel2024}. Considering the loss of the BlogForever project, we conclude that sustainability is unlikely to be achieved by constructing a dedicated blog repository in isolation from the existing infrastructure landscape, but rather by embedding blog preservation into services that are already operationally maintained. Additionally, distributing the responsibility for blog preservation and ensuring long-term access by multiple stakeholders reduces the risk of data loss \autocite{ochsner2026_requirements_preprint}.

\paragraph{Towards a cooperative infrastructure} Our conceptual synthesis supports the view that the long-term preservation of scholarly blogs will not emerge from a single institution. Rather, it requires the coordinated contributions of blogging platforms, dedicated blog archives, national libraries, subject-specific information services, and institutional repositories. Each of these actors already performs parts of the required functions, but responsibilities overlap in some areas and leave gaps in others \autocite{ochsner2025_jdoc, pampel2024}. The catalog can serve as a shared reference point against which these distributed activities can be aligned. Within this constellation, Rogue Scholar is a particularly promising candidate for an aggregator role, given its existing technical capacity for feed-based ingest, DOI assignment, Internet Archive coordination, and open licensing \autocite{fenner2023}. A formalized cooperation with national libraries and an integration with subject-specific and institutional repositories would be a logical and operationally feasible next step.

\paragraph{Focus on Germany} The catalog was developed with explicit reference to the German information infrastructure, which is characterized by a distinctly federal organization, a strong tradition of subject-specific provision through the FIDs, and a well-established division of labor between national, regional, and institutional actors. This focus is both a limitation and a methodological strength. It is a limitation insofar as the specific configuration of actors differs across national contexts and the catalog cannot be transferred one-to-one to other countries. It is a strength insofar as the German landscape is unusually heterogeneous and therefore provides a demanding test case for a cooperative model: requirements that can be formulated coherently across such a distributed landscape are likely to be applicable, with appropriate adaptation, to more centralized national systems as well. The categories of actors addressed in the catalog — scholarly blogging platforms, dedicated blog archives, national libraries, subject-specific information services, institutional repositories, and union catalogs — exist in analogous forms in many national information infrastructures, and the requirements are formulated at a level of abstraction that supports such transfer.

\paragraph{Theoretical implications} Beyond its practical relevance, the catalog contributes to the broader discussion about the scope and composition of the scholarly record in the Open Science era. Traditional definitions are oriented towards stable, editorially validated, and citable artifacts. Scholarly blogs challenge this definition along several axes: they are often informally edited, revised after publication, and distributed across heterogeneous platforms. Taking their preservation seriously therefore implies that the scholarly record is neither uniform in format nor singular in custody, but rather a distributed, versioned, and cooperatively maintained entity. The catalog operationalizes this view for one specific class of digital scholarly objects and may serve as a template for analogous efforts concerning preprints with continuous versioning, notebook-based publications, and research software.

\paragraph{Methodological reflection} The convergent mixed-methods foundation of this article \autocite{creswell2017} allowed us to combine the breadth of the quantitative mapping with the depth of the qualitative interviews and to translate both into conceptual recommendations. Triangulation across the two empirical strands increases the trustworthiness of our conceptual work, as, for example, the low prevalence of DOIs observed quantitatively is consistent with the interview reports of high barriers to DOI assignment. The open, participatory review process complemented these strands with perspectives beyond the authors and interview participants. We acknowledge, however, that the 22 written comments and the 40 workshop participants are not representative in a statistical sense and tend to self-select for engaged members of the scholarly blogging community. Their contributions should therefore be understood as a broadening, not as a validation, of the catalog.

\paragraph{Limitations and future research} The catalog is grounded in quantitative data describing the integration of blogs into information infrastructures and a qualitative sample of 13 bloggers that described their requirements for such an infrastructure. Within both samples blogs from the natural sciences, life sciences, and engineering sciences are under-represented, and requirements concerning content representation and references may need to be revised for fields with heavier reliance on code, data, or formalized notations. The catalog is additionally formulated at a level of abstraction that is appropriate for cross-institutional alignment but still leaves substantial room for interpretation at the level of concrete implementation. Future research should therefore (i) test the catalog through concrete implementation pilots with blogging platforms, subject-specific information services, institutional repositories, and national libraries, in order to identify practical obstacles and unanticipated side-effects; (ii) extend the empirical basis to additional disciplines and to blogs in languages other than German and English; (iii) investigate sustainable governance and funding models for a cooperative infrastructure, including the equitable distribution of costs for DOI registration, hosting, and long-term preservation; and (iv) compare the German situation with approaches taken in other countries, in order to inform the development of an international, interoperable ecosystem for the preservation of scholarly blogs.
\section{Conclusion}
Scholarly blogs are an established medium of scholarly communication, yet they remain only partially integrated into the information infrastructures that secure the long-term accessibility of the scholarly record. Prior work has quantitatively documented the gaps in this integration \autocite{ochsner2025_jdoc, ochsner2025_bfp} and has qualitatively articulated the requirements of scholarly bloggers for a preservation infrastructure \autocite{ochsner2026_requirements_preprint}. Building on this empirical foundation, the present article develops an implementation-oriented, conceptual catalog of requirements for a cooperative information infrastructure for scholarly blogs.

The catalog addresses four dimensions that jointly define such an infrastructure: functional requirements concerning persistent identifiers, metadata, repository integration, content representation, references, versioning, and searchability (F-01 to F-07); requirements to ensure data protection, in particular GDPR compliance and data sovereignty (D-01, D-02); requirements for the aggregation of blogs, namely an opt-in model and active guidance and support for bloggers and institutions (A-01, A-02); and requirements for interfaces governing the import and export of blog content to and from repositories and discovery services (I-01, I-02). Together, these thirteen requirements describe an infrastructure that is grounded in established technical standards, organizationally distributed across existing actors, and designed to respect the agency of scholarly bloggers. Our central argument is that the long-term preservation of scholarly blogs should be understood as a cooperative task. No single institution can or should provide the full set of services described in the catalog. 

Web archives, dedicated blog archives, national libraries, subject-specific information services, institutional repositories, union catalogs and periodicals databases, and scholarly blogging platform providers each already perform parts of the required functions — in the German context, for example, de.hypotheses, Rogue Scholar, the German National Library, the Specialized Information Services, and the institutional repositories of universities and research performing organizations. The catalog provides a shared reference against which these distributed activities and actors can be aligned and through which gaps and overlaps can be identified. Within such a constellation, dedicated blog archives such as Rogue Scholar emerge as a promising candidate for the role of an aggregator that mediates between blogs, DOI registration agencies, web archives, and memory institutions.

The contributions of this article are therefore both practical and theoretical. Practically, the catalog offers a structured and instructive tool for scholarly bloggers who wish to make their blogs more sustainable, for blogging platform providers to support preservation workflows, and for libraries, archives, and repositories that seek to extend their mission to alternative forms of scholarly communication and scholarly output. Theoretically, the catalog operationalizes a view of the scholarly record as a distributed, versioned, and cooperatively maintained entity rather than as a singular, uniformly custodied artefact. Furthermore, the catalog may inform analogous efforts for other alternative formats such as preprints, notebook-based publications, and research software.

The next step is implementation. We invite interested stakeholders — blogging platforms, libraries and archives along with their institutional repositories, and scholarly bloggers themselves — to test the catalog in concrete preservation pilots, to report on practical obstacles, and to contribute to its ongoing refinement. Only through such shared implementation will the catalog fulfill its purpose: to ensure that the scholarly blogosphere is preserved not as a fragmented collection of disappearing websites, but as an integral and lasting part of the scholarly record.

\section{Funding}
The work of Catharina Ochsner is funded by the German Research Foundation (DFG) through the project Kooperative Informationsinfrastruktur für wissenschaftliche Blogs (Infra Wiss Blogs) (project number 528958385). Heinz Pampel was partly funded by the Einstein Center Digital Future (ECDF).

\newpage
\printbibliography[heading=bibintoc]

\newpage
\renewcommand{\thepage}{\Alph{page}}
\setcounter{page}{1}

\end{document}